# Reduced perplexity:
# A simplified perspective on assessing probabilistic forecasts

Kenric P. Nelson[1,2]

*A simple, intuitive approach to the assessment of probabilistic inferences is introduced. The Shannon information metrics are translated to the probability domain. The translation shows that the negative logarithmic score and the geometric mean are equivalent measures of the accuracy of a probabilistic inference. Thus there is both a quantitative reduction in perplexity, which is the inverse of the geometric mean of the probabilities, as good inference algorithms reduce the uncertainty and a qualitative reduction due to the increased clarity between the original set of probabilistic forecasts and their central tendency, the geometric mean. Further insight is provided by showing that the Rényi and Tsallis entropy functions translated to the probability domain are both the weighted generalized mean of the distribution. The generalized mean of probabilistic forecasts forms a spectrum of performance metrics referred to as a Risk Profile. The arithmetic mean is used to measure the decisiveness, while the -2/3 mean is used to measure the robustness.*

## 1    Introduction

The objective of this chapter is to introduce a clear and simple approach to assessing the performance of probabilistic forecasts. This is important because machine learning and other techniques for decision making are often only evaluated in terms of percentage of correct decisions. Management of uncertainty in these systems requires that accurate probabilities be assigned to decisions. Unfortunately the existing assessment methods based on "scoring rules" [1], [2], which are defined later in the chapter, are poorly understood and often misapplied and/or misinterpreted. The approach here will be to ground the assessment of probability forecasts using information theory [3]–[5], while framing the results from the perspective of the central tendency and fluctuation of the forecasted probabilities. The methods will be shown to reduce both the colloquial perplexity surrounding how to evaluate inferences and the quantitative perplexity which is an information-theoretic measure related to the accuracy of the probabilities.

To achieve this objective, Section 2 reviews the relationship between probabilities, perplexity and entropy. The geometric mean of probabilities is shown to be the central tendency of a set of probabilities. In Section 3, the relationship between probabilities and entropy is expanded to include generalized entropy functions [6], [7]. From this, the generalized mean of probabilities is shown to provide insight into the fluctuations and risk-sensitivity of a forecast. From this analysis, a *Risk Profile* [8] defined as the spectrum of generalized means of a set of forecasted probabilities, is used in Section 4 to evaluate a variety of models for a *n*-dimensional random variable.

[1] Raytheon Company, 235 Presidential Way, Woburn, MA 01801, kenric_p_nelson@raytheon.com
[2] Boston University, 8 St Mary's St, Boston, MA 02215, kenricpn@bu.edu

## 2   Probability, Perplexity and Entropy

The arithmetic mean and the standard deviation of a distribution are the elementary statistics used to describe the central tendency and uncertainty respectively of a random variable. Less widely understood, though studied as early as the 1870s by McCalister [9], is that a random variable which is formed by the ratio of two independent random variable has a central tendency determined by the geometric mean rather than the arithmetic mean. Thus the central tendency of a set of probabilities, each of which is formed from a ratio, is determined by their geometric mean. This property will be derived from information theory and illustrated with the example of the Gaussian distribution.

Instead of using the geometric mean of the probabilities of a distribution to represent average uncertainty, long-standing tradition within mathematical physics has been to utilize the entropy function, which is defined using the arithmetic mean of the logarithm of the probabilities. There are at least three important reasons for this. Physically, entropy defines the change in heat energy per temperature; mathematically, entropy provides an additive scale for measuring uncertainty; and computationally, entropy has been shown to be a measure of information [10], [11]. Unfortunately, in using entropy to quantify average uncertainty, what is lost is the intuitive relationship between the underlying probabilities of a distribution and a summarizing average probability of the distribution. Perplexity, which determines the average number of uncertain states, provides a bridge between the average probability and the entropy of a distribution. For a random variable with a uniform distribution of $N$ states, the perplexity is $N$ and its inverse $1/N$ is the average probability. More generally, the average probability $P_{avg}$ and the perplexity $PP$ are related to the entropy $H(\mathbf{p}) = -\sum_{i=1}^{N} p_i \ln p_i$ of a distribution $\mathbf{p} = \left\{ p_i : \sum_{i=1}^{N} p_i = 1 \right\}$ by

$$P_{avg} \equiv PP^{-1} = \exp(-H(\mathbf{p})) = \exp\left( \sum_{i=1}^{N} p_i \ln p_i \right) = \prod_{i=1}^{N} p_i^{p_i} . \tag{2.1}$$

The expression on the far right is the weighted geometric mean of the probabilities in which the weight appearing in the exponent is also the probability. For a continuous distribution $f(x)$ of a random variable $X$ these expressions become

$$f_{avg} \equiv PP^{-1} = \exp(-H(f(x))) = \exp\left( \int_{x \in X} f(x) \ln f(x) dx \right), \tag{2.2}$$

where $f_{avg}$ is the average density of the distribution and *PP* still refers to perplexity. The term on the right cannot be simplified further, since there is not a continuous form of the product function.

Figure 1 illustrates these relationships for the standard normal distribution. The key point is that by expressing the central tendency of a distribution as a probability (or density for continuous distributions), the context with the original distribution is maintained. For the exponential and Gaussian distributions, translating entropy back to the density domain [12] results in the density of the distribution at the location $\mu$ plus the scale $\sigma$

$$\exp\left(\int_{\mu}^{\infty} \frac{1}{\sigma} e^{-\frac{x-\mu}{\sigma}} \ln\left(\frac{1}{\sigma} e^{-\frac{x-\mu}{\sigma}}\right) dx\right) = \frac{1}{\sigma} e^{-\left(\frac{\mu+\sigma-\mu}{\sigma}\right)} = \frac{1}{\sigma e}, \qquad (2.3)$$

$$\exp\left(\int_{-\infty}^{\infty} \frac{1}{\sqrt{2\pi}\sigma} e^{-\frac{1}{2}\left(\frac{x-\mu}{\sigma}\right)^2} \ln\left(\frac{1}{\sqrt{2\pi}\sigma} e^{-\frac{1}{2}\left(\frac{x-\mu}{\sigma}\right)^2}\right) dx\right) = \frac{1}{\sqrt{2\pi}\sigma} e^{-\frac{1}{2}\left(\frac{x-\mu}{\sigma}\right)^2} = \frac{1}{\sqrt{2\pi e}\sigma}. \qquad (2.4)$$

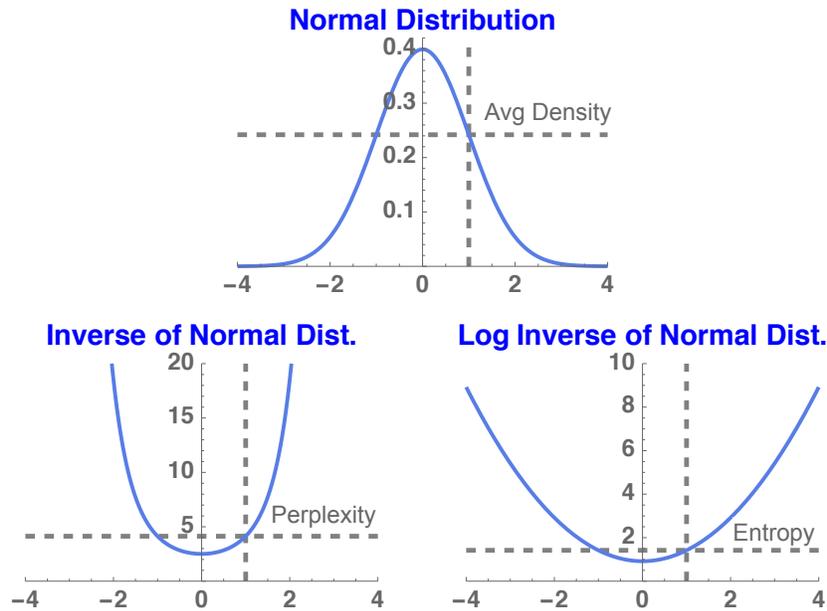

**Figure 1** Comparison of the average density, perplexity and entropy for the standard normal distribution. Plots of the inverse distribution and the log of the inverse of the distribution provide visualization of the perplexity and entropy. The intersection for each of these quantities with the distribution is at the mean plus the standard deviation.

Thus, it should be more commonly understood that for these two important members of the exponential family, the average uncertainty is the density at the width of the distribution defined by $f(\mu+\sigma)$. While perplexity and entropy are valuable concepts, it is not common to plot distributions on the inverse scale (perplexity) or the log inverse scale (entropy), thus the intuitive meaning of these quantities is disconnected from the underlying distribution.

Table 1 shows the translation of entropy, divergence and cross-entropy to the perplexity and probability scales. In each case the translation is via application of the exponential function as in (2.1). The additive combination of logarithmic probabilities, translates into a multiplicative combination of the probabilities. The weight on the mean, also a probability, is now a power term. The additive relationship between cross-entropy, entropy, and divergence $H(p,q) = H(p) + D_{KL}(p||q)$, is multiplicative in the probability space

$$P_{cross-entropy} = P_{entropy} P_{divergence}$$
$$= \left( \prod_{i=1}^{N} p_i^{p_i} \right) \left( \prod_{i=1}^{N} \left( q_i / p_i \right)^{p_i} \right) \quad (2.5)$$
$$= \prod_{i=1}^{N} q_i^{p_i}.$$

Jaynes [13], [14] established the principal of maximum entropy as a method for selecting a probability distribution such that known constraints were satisfied, but no additional knowledge was represented in the distribution. Two basic examples are the exponential distribution, which satisfies the constraint that the range is 0 to ∞ and a known mean $\mathbb{E}(X) = \int_0^\infty x f(x) dx = \mu$, and the Gaussian distribution which satisfies a known mean and variance $\mathbb{E}(X^2) - \mathbb{E}(X)^2 = \int_{-\infty}^{\infty} (x-\mu)^2 f(x) dx = \sigma^2$. Translated to the probability domain, the principal of maximum entropy can thus be framed as a minimization of the weighted geometric mean of the distribution. In section 4 a related principal of minimizing the cross entropy between a discrimination model and the actual uncertainty of a forecasted random event will be translated to maximizing the geometric mean of the reported probability.

**Table 1:** Translation of entropy functions to perplexity and probability scales

| Info-Metric | Entropy Scale | Perplexity Scale | Probability Scale |
|---|---|---|---|
| Entropy | $-\sum_i p_i \ln p_i$ | $\prod_i (p_i)^{-p_i}$ | $\prod_i (p_i)^{p_i}$ |
| Divergence | $-\sum_i p_i \ln \left( q_i / p_i \right)$ | $\prod_i \left( q_i / p_i \right)^{-p_i}$ | $\prod_i \left( q_i / p_i \right)^{p_i}$ |
| Cross-Entropy | $-\sum_i p_i \ln q_i$ | $\prod_i (q_i)^{-p_i}$ | $\prod_i (q_i)^{p_i}$ |

Just as the arithmetic mean of the logarithm of a probability distribution determines the central tendency of the uncertainty or the entropy, the standard deviation of the logarithm of the probabilities, $\sigma_{\ln p}$, is needed to quantify variations in the uncertainty,

$$\sigma_{\ln p} \equiv \left[ \sum_{i=1}^{N} p_i \left(-\ln p_i\right)^2 - \left(-\sum_{i=1}^{N} p_i \ln p_i\right)^2 \right]^{1/2}. \tag{2.6}$$

Unfortunately, the translation to the probability domain ($e^{-\sigma_{\ln p}}$) does not result in a simple function with a clear interpretation. Furthermore, because the domain of entropy is one-sided, just determining the standard deviation does not capture the asymmetry in the distribution of the logarithm of the probabilities. In the next section, the generalized mean of the probabilities is shown to be a better alternative for measuring fluctuations.

## 3   Relationship between the generalized entropy and the generalized mean

In this section, the effect of sensitivity to risk ($r$) will be used to generalize the assessment of probabilistic forecasts. The approach is based on a generalization of the entropy function, particularly the Rényi and Tsallis entropies [7], [15], [16]. As with the Boltzmann-Gibbs-Shannon entropy, the generalized entropy can be transformed back to the probability domain. The resulting function, derived in [12] and summarized in the Appendix, is the weighted generalized mean or weighted $p$-norm of the probabilities

$$P_r(\mathbf{w}, \mathbf{p}) \equiv \begin{cases} \left(\sum_{i=1}^{N} w_i p_i^r\right) & r \neq 0 \\ \prod_{i=1}^{N} p_i^{w_i} & r = 0, \end{cases} \tag{2.6}$$

where the symbol $r$ is used for the power of the mean to avoid confusion with the probabilities and because the power is related to the sensitivity to risk, as discussed below. The weights $\mathbf{w} = \left\{ w_i : \sum_{i=1}^{N} w_i = 1 \right\}$ are a modified version of the probabilities discussed in the next paragraph. The symbol $P_r$ is used here rather than the traditional symbols of $M_r$ or $\|x\|_r$ for the generalized mean and $p$-norm respectively to emphasize that the result is a probability which represents a particular aggregation of the vector of probabilities. The geometric mean, which is the metric consistent with Shannon information theory, is recovered when the risk sensitivity is zero ($r = 0$). Positive values of $r$ reduce the influence of low probabilities in the average and are thus associated with risk-seeking, while negative values of $r$ increase the sensitivity to low probabilities and are thus risk-averse.

Several different generalizations of entropy can be shown to transform into the form of (2.6). The Appendix discusses the origin of these generalizations for modeling the statistical properties of complex systems which are influenced by nonlinear coupling. The Tsallis and normalized Tsallis entropy utilize a modified set of probabilities formed by raising the probabilities of the distribution to a power and renormalizing

$$P_i^{(r)}(\mathbf{p}) \equiv \frac{p_i^{1-r}}{\sum_{j=1}^{N} p_j^{1-r}}. \tag{2.7}$$

This new distribution, referred to either as coupled probability or an escort probability, is the normalized probability of $1-r$ independent events rather than one event. Substituting (2.7) for the weights in (2.6) and simplifying gives the following expression for the weighted generalized mean of a distribution

$$P_r(\mathbf{P}^{(r)}, \mathbf{p}) = \left( \sum_{i=1}^{N} \left( \frac{p_i^{1-r}}{\sum_{j=1}^{N} p_j^{1-r}} \right) p_i^r \right)^{\frac{1}{r}} = \left( \sum_{i=1}^{N} p_i^{1-r} \right)^{\frac{-1}{r}} = P_{-r}(\mathbf{p},\mathbf{p}). \tag{2.8}$$

The normalized probability of $1-r$ events as a weight has the effect of reversing the sign of power $r$ with the original probabilities now the weights, as shown in the right most expression.

Figure 2 which shows the weighted geometric mean for three different distributions is plotted in terms of $-r$ so the visual orientation of graphs is similar to those appearing later regarding assessment of probabilistic forecasts. The distributions examined are members of the coupled-Gaussians, which are equivalent to the Student's-t and discussed in more detail in the Appendix. For consistency, the coupled-Gaussians distributions are expressed here in terms of the risk sensitivity $r_D$ and the subscript $D$ is used to distinguish the parameter of the distribution from the parameter of the generalized mean. The coupled-Gaussian is

$$f(x) = \frac{1}{Z(r_D,\sigma)} \left( 1 - \left( \frac{r_D}{2+r_D} \right) \frac{x^2}{\sigma^2} \right)_+^{\frac{1}{r_D}}, \tag{2.8}$$

where $(a)_+ \equiv \max(0,a)$, $Z$ is the normalization of the distribution and $\sigma$ is the scale parameter of the distribution. For $-2 < r_D < 0$ the distribution is heavy tail, $r_D = 0$ is the Gaussian, and $r_D > 0$ is a compact-support distribution. Applying the continuous form of the generalized mean (2.8) with the matching power of $r = r_D$ gives the following result

$$f_{r_D}(f(x,r_D,\sigma)) = \left( \int_{x \in X} f(x,r_D,\sigma)^{1-r_D} dx \right)^{\frac{1}{-r_D}}$$

$$= Z(r_D,\sigma)^{\frac{1-r_D}{r_D}} \left( \int_{x \in X} \left( 1 - \left( \frac{r_D}{2+r_D} \right) \frac{x^2}{\sigma^2} \right)^{\frac{1-r_D}{r_D}} dx \right)_+^{\frac{1}{-r_D}}$$

$$= Z(r_D,\sigma)^{\frac{1-r_D}{r_D}} \left( Z(r_D,\sigma) \left( 1 - \frac{r_D}{2+r_D} \right)^{-1} \right)^{\frac{1}{-r_D}}$$

$$= \frac{1}{Z(r_D,\sigma)} \left( 1 - \frac{r_D}{2+r_D} \right)^{\frac{1}{r_D}} = f(x = \sigma, r_D, \sigma).$$

(2.9)

That is, the generalized mean of the coupled Gaussian with a matching risk sensitivity is equal to the density at the mean plus the scale. While not derived here, the equivalence between of the generalized maximum entropy principal using the Tsallis entropy and the minimization of the weighted generalized mean is such that the distribution $f(x, r_D, \sigma)$ is the minimization of $f_{r_D}$ given the constraint that the scale is σ.

In Figure 2 the weighted generalized mean (wgm) is shown for the Gaussian distribution $r_D = 0$ and two examples of the Coupled Gaussian with $r_D = -2/3, 1$. As derived in the appendix, these values of risk sensitivity are conjugate values in the heavy-tail and compact-support domain, respectively. For each of the distributions the scale is $\sigma = 1$. In order to illustrate the intersection between the distribution and its matching value of the wgm, the mean of each distribution is shifted by $\mu = r_D - \sigma$. The wgm is plotted as a function of $2r_D - r$ rather than $r$ so that the increase in wgm is from left to right, as it will be when evaluating probabilistic forecasts. The coupled exponential distribution and the coupled Gaussian distribution have the following relationship with respect to the generalized average uncertainty

$$\exp_r \left( \int_\mu^\infty \frac{1}{\sigma} \exp_r \left( -\frac{x-\mu}{\sigma} \right) \ln_r \left( \frac{1}{\sigma} \exp_r \left( -\frac{x-\mu}{\sigma} \right) \right) dx \right) = \frac{e_r^{-\left(\frac{\mu+\sigma-\mu}{\sigma}\right)}}{\sigma} = \frac{1}{\sigma e_r}$$

(2.9)

$$\exp_r \left( \int_{-\infty}^\infty \frac{1}{Z_r} e^{-\frac{(x-\mu)^2}{2\sigma^2}} \ln_r \left( \frac{1}{Z_r} e^{-\frac{(x-\mu)^2}{2\sigma^2}} \right) dx \right) = \frac{1}{Z_r} e_r^{-\frac{(\mu+\sigma-\mu)^2}{2\sigma^2}} = \frac{e_r^{-1/2}}{Z_r},$$

(2.9)

where the subscript *D* was dropped for readability. These relationships provide evidence of the importance of the generalized mean as an expression of the average uncertainty for non-exponential distributions. In the next section, use of the generalized mean as a metric to evaluate probabilistic inference is demonstrated.

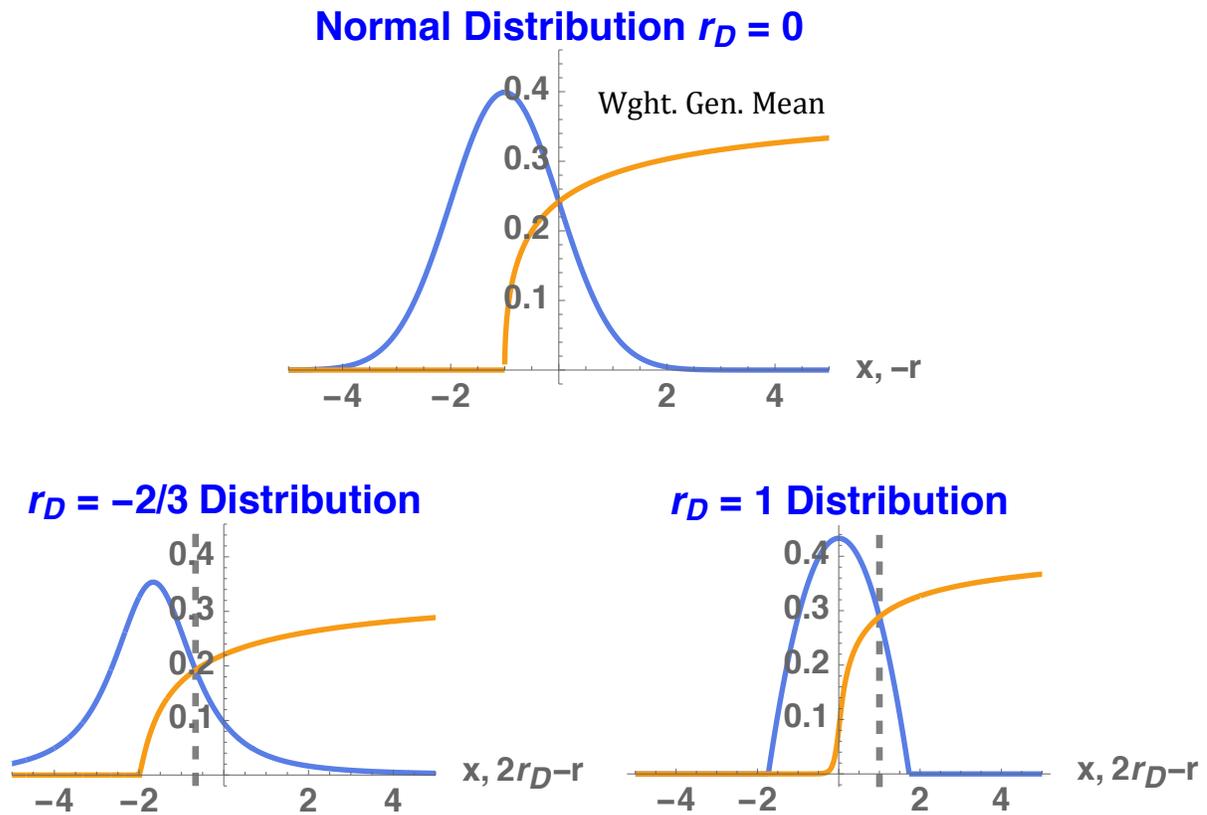

**Figure 2** Plots of the weighted generalized mean (wgm) overlayed with the distribution which minimizes the wgm at the value $r = r_D$. The mean of each distribution is adjusted to show the wgm intersecting the density at the mean plus width parameter of the distribution. a) Normal distribution N(-1,1) (blue) with its wgm (orange). The normal distribution is a coupled Gaussian with $r_D = 0$ and minimizes the wgm at $r = 0$ (weighted geometric mean). The wgm at $r = 0$ is equal to the density of the normal at the mean plus standard deviation. b) The coupled Gaussian with $r_D = -2/3$, $\mu = -5/3$, $\sigma = 1$ minimizes the wgm at $r = -2/3$. The orientation of the wgm plot for b and c is inverted and shifted by $2r_D - r$. c) The coupled Gaussian with $r_D = 1$, $\mu = 0$, $\sigma = 1$ minimizes the wgm at $r = 1$. For both b and c the wgm at $r_D$ is equal to the density at the mean plus the generalized standard deviation.

## 4 Assessing probabilistic forecasts using a Risk Profile

The goal of an effective probabilistic forecast is to "reduce perplexity"; i.e. to enhance decision making by providing accurate information about the underlying uncertainties. Just as the maximum entropy approach is important in selecting a model which properly expresses the uncertainty, minimization of the cross-entropy between a model and a source of data is essential to accurate forecasting. In Section 3 the relationship between the weighted generalized mean of a distribution and the generalized entropy functions was established; likewise, the generalized cross-entropy can be translated into the weighted generalized mean in probability space. The result is a spectrum of metrics which modifies the sensitivity to surprising or low-probability events; as such it is referred to as a *Risk Profile*.

The most basic definition of risk $R$ is the expected cost of a loss $L$ times the probability of the loss

$$R = E(L) = \sum_{i=1}^{N} L_i p(L_i). \tag{3.1}$$

The risk can also be defined as the degree of variance or standard-deviation for a process, such as an asset price, which has a monetary or more general value. An individual or agent can have different perceptions of risk, expressed as the utility of a loss (or gain). Thus a risk-averse person would seek to lower exposure to high variances given the same expected loss. With regard to a probabilistic forecast, the cost is being surprised by an event which was forecasted to have a low probability. While a particular application may also assign a valuation to events, with regard to evaluating the quality of the forecast itself, the 'surprisal ($S$)' will be the only cost. A neutral perspective on the risk of being surprised is the information theoretic measure, the logarithm of the probabilities [3], [17]. The expected surprisal cost is the arithmetic average of the logarithmic distance between the forecasted probabilities and a perfect forecast of $p=1$

$$S = E[S_i] = -\tfrac{1}{N}\sum_{i=1}^{N}(\ln p_i - \ln 1) = -\tfrac{1}{N}\sum_{i=1}^{N}\ln p_i. \tag{3.1}$$

The average surprisal is also known as the logarithmic scoring rule or the negative log-likelihood of the forecasts and has the property of being the only scoring rule which is both proper and local. A proper scoring rule is one in which optimization of the rule leads to unbiased forecasts relative to what is known by the forecaster. A local scoring rule is one in which only the probabilities of events which occurred are used in the evaluation. The average surprisal (3.1) can be viewed as the cross-entropy between a model (the reported probabilities) and data (the distribution of the test set). From (2.5) the uncertainty in a forecast is due to underlying uncertainty in the test set (entropy) and errors in the model (divergence). This relationship is used in [18] to visualize the quantitative performance of forecasts along side the calibration curve comparing the forecasted and actual distribution.

The influence of risk-seeking and risk-aversion in forecasting can be evaluated using a generalized surprisal function which is defined as

$$S_r \equiv E[S_{r,i}] = -\tfrac{1}{N}\sum_{i=1}^{N}(\ln_r p_i - \ln_r 1) = -\sum_{i=1}^{N}\ln_r p_i$$

$$\ln_r x \equiv \frac{1+r}{r}(x^r - 1). \tag{3.2}$$

This generalized logarithmic function is fundamental to the generalization of thermodynamics introduced by Tsallis [7]. Its role in defining a generalized information theory is explained further in the Appendix. The generalized surprisal function is still a local scoring rule, but is no longer proper. The properties of this function have been

Nelson, Reduced Perplexity                                                                                                      P a g e | 9

studied in economics due to its preservation of a constant coefficient of relative risk. In economics the variable *x* of (3.2) is the valuation and the relative risk aversion [19], [20] is defined in terms of $1-r$ since $r=1$ is a linear function and thus is considered to be neutral risk. Here the bias is with respect to the neutral measure of information, namely $\ln p$ when $r=0$. Thus for purposes of this discussion the relative risk sensitivity is defined as

$$r \equiv 1 + p \frac{d^2(\ln_r p)/dp^2}{d(\ln_r p)/dp}. \tag{3.3}$$

For negative values of *r*, the generalized surprisal is risk-averse, since the cost of being surprised goes to infinity faster. This is referred to as the domain of robust metrics, since it encourages algorithms to be conservative or robust in probabilistic estimation. For positive values of *r*, the measure is risk-seeking and is referred to as a decisive metric since it is more like the cost of making a decision over a finite set of choices, as opposed to the cost of properly forecasting the probability of the decision.

For evaluating a probabilistic forecast, use of the logarithmic or generalized logarithmic scale is needed to assure that the analysis properly measures the cost of a surprising forecast; nevertheless, it leaves obscure what is truly desired in an evaluation: knowledge of the central tendency and fluctuation of the probability forecasts. Following the procedures introduced in Sections 2 and 3, the generalized scoring rule can be translated to a probability by taking the inverse of the generalized logarithm, which is the generalized exponential

$$\exp_r(x) \equiv \begin{cases} \left(1 + \frac{r}{1+r}x\right)_+^{\frac{1}{r}} & r \neq 0 \\ \exp(x) & r = 0, \end{cases} \tag{3.4}$$

where $(a)_+ \equiv \max(0,a)$. Applying (3.4) to (3.2) shows that the generalized mean of the probabilities is the translation of the generalized logarithmic scoring rule to the probability domain

$$P_{r-avg}(\boldsymbol{p}) \equiv \exp_r(-S_r(\boldsymbol{p})) = \left(1 + \frac{r}{1+r}\left(\frac{1}{N}\sum_{i=1}^{N}\frac{1+r}{r}(p_i^r - 1)\right)\right)^{\frac{1}{r}} = \begin{cases} \left(\frac{1}{N}\sum_{i=1}^{N} p_i^r\right)^{\frac{1}{r}} & r \neq 0 \\ \prod_{i=1}^{N} p_i^{1/N} & r = 0. \end{cases} \tag{3.5}$$

Thus, the generalized mean of the forecasted probabilities forms a spectrum of metrics which profile the performance of the forecast relative to the degree of relative risk sensitivity. This spectrum is referred to as the *Risk Profile* of the probabilistic forecast. A condensed summary of an algorithms performance is achieved using three points on the spectrum: the geometric mean ($r=0$) measures the risk-neutral *accuracy* and the degree of fluctuation is measured by an upper metric called *decisive* using the arithmetic mean (*r=1*) and a lower metric called *robustness* using the $-2/3^{\text{rds}}$ mean ($r=-2/3$).

Prior to demonstrating the Risk Profile, a word of caution regarding the use of proper scoring rules, such as the mean-square average, is provided.  Starting with Brier [21], a tradition has grown around the use of non-information theoretic measures of accuracy for probabilistic forecasts.  Brier himself cannot be faulted, as Shannon's efforts to formulate information theory [3] were nearly concurrent with Brier's efforts to evaluate weather forecasts.  However, the subsequent development of proper scoring rules [2], [22], [23], which removes the bias in the expectation of a forecast optimized using any convex positive-valued utility function, has led to the impression that an unbiased expectation is the only criteria for evaluating forecasts.  While some applications may in fact require a utility function different from information theory, in practice use of the mean-square average because it is "proper" has inappropriately justified avoidance of the rigorous information-theoretic penalties for over-confident forecasts.  One way to view this is that while the first-order expectation is unbiased for a proper scoring rule, all the other moments of the forecasts may still be biased.  A rigorous proof of this deficiency would be a valuable contribution as suggested by Jewson [24].  Here, the emphasis is on using the alternative cost functions to complement the Shannon info-metric and thereby to provide insight into how an algorithm responds to risk sensitivity.  As derived by Musio and Dawid [25], the generalized surprisal with $r=1$, used here for a measure of decisiveness, becomes the mean-square average scoring rule if the distance between the non-event forecasts and a probability of zero is included to make a proper score.

To illustrate the Risk Profile in evaluating statistical models, the contrast between robust and decisive models of a multivariate Gaussian random variable is demonstrated. The Student's t-distribution originated from the insight by William Gosset [26] that a limited number of samples from a source known to have a Gaussian distribution requires a model which modifies the Gaussian distribution to have a slower than exponential decay. Again, using the equivalent coupled-Gaussian distribution (2.8), but now for a multi-dimensional variable the distribution is

$$G_{r_D}(\mathbf{x};\,\boldsymbol{\mu},\boldsymbol{\Sigma}) \equiv \frac{1}{Z_{r_D}(\boldsymbol{\Sigma})}\left(1 - \frac{r_D}{2+r_D}(\mathbf{x}-\boldsymbol{\mu})^{\text{T}} \cdot \boldsymbol{\Sigma}^{-1} \cdot (\mathbf{x}-\boldsymbol{\mu})\right)_{+}^{\frac{1}{r_D}}, \qquad (3.6)$$

where the vectors **x** and **μ** are the random variable and mean vectors, **Σ** is the correlation matrix[3], and $Z_r$ is the normalization.  The parameter $r_D$ has a dependence on the dimension which is explained in the Appendix.

---

[3] While **Σ** is the covariance matrix for a Gaussian distribution, for the coupled-Gaussian this matrix is a generalization of the covariance and like the Student's t distribution is known as the correlation matrix.

The problems with trying to model a Gaussian random variable using a multivariate Gaussian as the model is shown in Figure 3. In this example 10 independent features, which are generated from Gaussian distributions, are modelled as an multivariate Gaussian with a varying number of dimensions based on estimates of the mean and standard deviation from 25 samples. Although reasonable classification performance is achieved (84%), the accuracy of the modelled probabilities is reduced beyond 6 dimensions. Furthermore, the robustness as measured by the -2/3 generalized mean drops to zero when all 10 dimensions are modelled.

Even without seeking to optimize the coupling value, improvement in the accuracy and robustness of the multivariate model can be achieved using heavy-tail decay. Figure 4 shows an example with $r_D = -0.15$ in which the accuracy is improved to 0.69 and is stable for dimensions 6-10. The robustness continues to decrease as the number of dimensions increases, but is improved significantly over the multivariate Gaussian model. The classification improves modestly to 86%, but is not the principal reason for using the heavy-tail model.

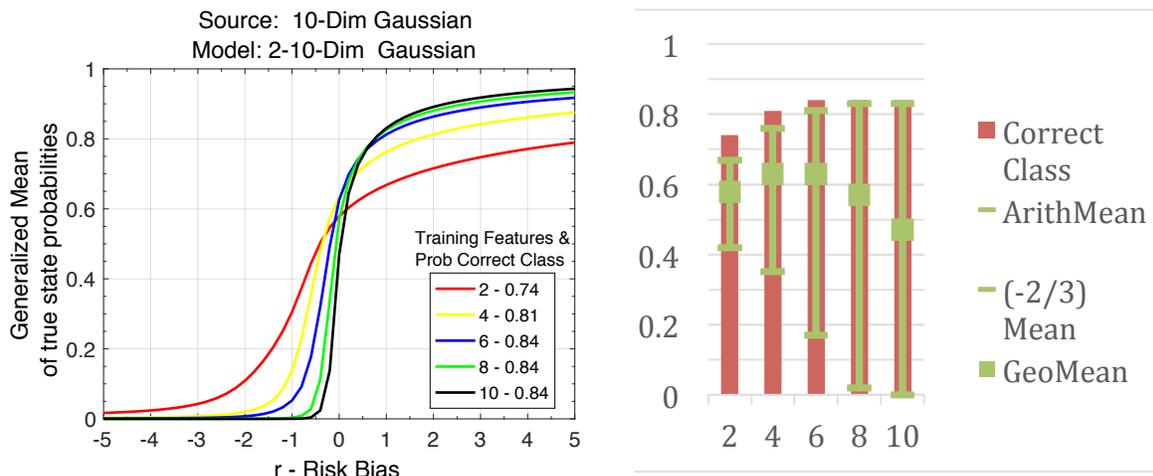

**Figure 3** A source of 10 independent Gaussian random variables is over-fit given 25 samples to learn the mean and variance of each dimension and a model which is also a multivariate Gaussian. a) The risk profile shows that as the number of dimensions increases the model becomes more decisive. b) At 6 dimensions, the classification performance saturates to 84% at and the accuracy of the probabilities reaches is maximum of 63%.

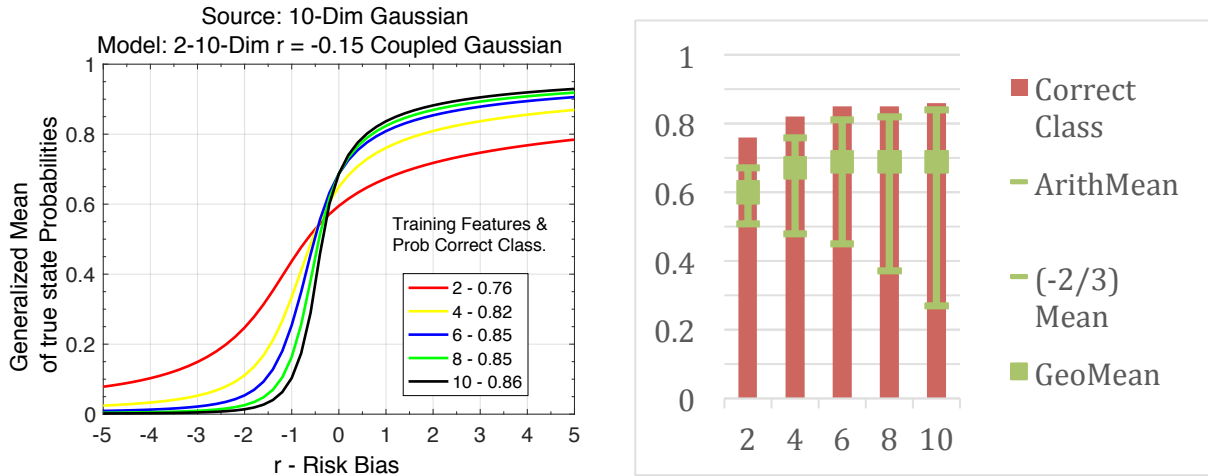

**Figure 4** The risk of overfitting is reduced by using a heavy-tail coupled-Gaussian. Shown is an example with $r = -0.15$. a) The risk profile shows that the accuracy of 0.69 continues to hold as the dimensions modeled is increased from 6 to 8. b) The percent correct classification (red bar) improves to 86% with 10 dimensions modeled. The robustness does go down as the number of dimensions is increased, but could be improved by optimizing the coupling value used.

The problems with overconfidence in the tails of a model are very visible when a compact-support distribution is used to model a source of data which is Gaussian. In this case, the reporting of $p = 0$ for states which do occur results in the accuracy being zero. An example of this situation is shown in Figure 5 in which the distribution power is $r_D = 0.6$. Although the model is neither accurate nor robust in the reporting of the probability of events, it is still capable of modest classification performance (75% for 4-dimensions and reduced to 67% for 10-dimensions); nevertheless, characterization of only the classification performance would not show the severity of the problem with inappropriately using a compact-support model.

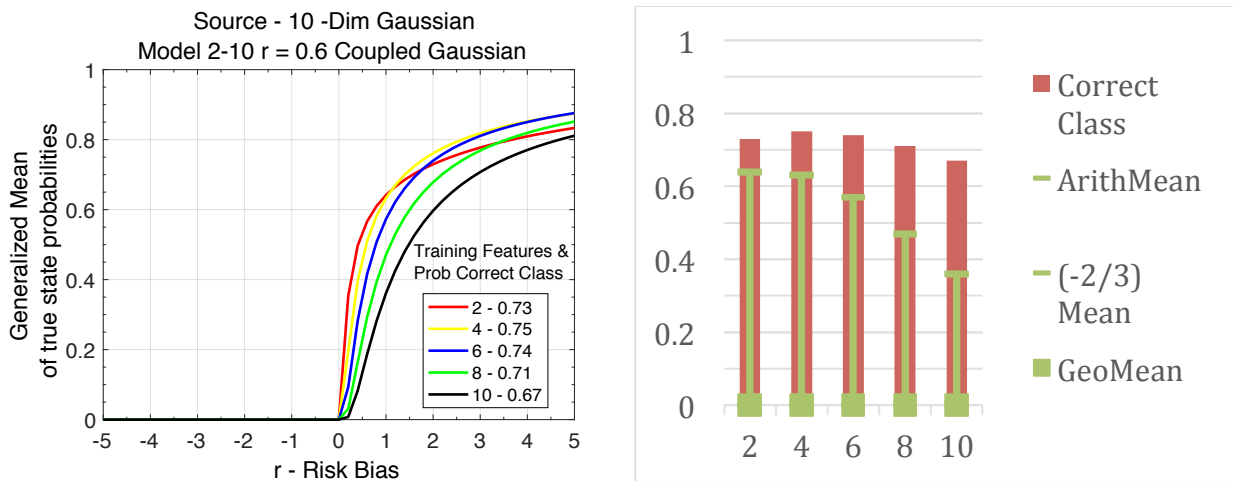

**Figure 5** Using a compact-support distribution to model a source of data which is Gaussian results in the probability accuracy being zero. a) The risk profile shows that the model using $r = 0.6$ is neither accurate nor robust. b) The classification performance (red bar) is only 75% for the model with four dimensions, but characterization of only the classification performance would not show the severity of the problem with this model.

# 5 Discussion and Conclusion

The purpose of this chapter has been to show that translating results of information theory from the entropy domain to the probability domain can simplify and clarify interpretation of important information metrics. In particular, the basic fact that the entropy of a Gaussian distribution when translated to a density as shown in (2.3) is equal to the density of the Gaussian at the mean plus the standard deviation should be a widely understood representation of the relationship between the standard deviation and entropy in measuring the central tendency of uncertainty. Unfortunately, while entropy provides the convenience of an additive information measure, the connection to the underlying probabilities of a distribution is often lost.

This disconnect between theory and practical intuition is evident in the confusion associated with evaluating probabilistic forecasts. For most random variables the 'average' is simply the arithmetic mean. Unfortunately, this does not hold for random variables which are formed by ratios, of which probabilities are a particularly important example. An elementary principal of probability theory is that the total probability of a set of independent probabilities is their product. So why isn't the $n^{th}$ root of the total probability or the geometric mean of the independent set of probabilities also recognized and taught to be the average? Likewise for a distribution, why isn't the weighted geometric mean in which the weight is also the probability recognized as the central tendency of the distribution. The answer seems to be both the misconception that the arithmetic mean is always the central tendency and the role that entropy serves in translating the geometric mean of probabilities to a domain in which the arithmetic mean is the central tendency.

For the evaluation of probabilistic forecasts, this has created a serious problem in which a variety of different 'proper scoring rules' are treated has having equal merit in assessing the central tendency of a forecasters performance. Only the logarithmic score, which is both proper (unbiased expectation) and local (based on actual events), is sensitive to the accuracy of the full distribution of forecasts. In particular, the Brier or mean-square average, which is a popular alternative, discounts the distance between small probabilities approaching zero. Thus although the average forecast may be unbiased when optimized using the mean-square average, the distribution of forecasts tends to be over-confident. A clear example is the allowance of a forecast of impossibility, i.e. a reported probability of zero, for events which actually do occur.

The perspective emphasized here is to use the logarithmic score or equivalently the geometric mean of the reported probabilities to measure the accuracy of forecasts. The biased scores or equivalently the generalized mean of the probabilities is used to measure the fluctuation of the forecasts. The generalized mean of the probabilities is derived from a generalized information theory which for decision-making models the sensitivity to risk. Rather than making these biased scores proper, their local property is maintained and they provide a *Risk Profile* which is sensitive to whether the forecasts tend to be under or over-confident. The arithmetic mean is biased toward decisive forecasting and approximates the classification performance over a small number of decisions. In contrast, means with a negative power are sensitive to the accuracy of rare events and thus provide a measure of the robustness of algorithms. The -2/3rds mean is conjugate to the arithmetic mean and is thus recommended as a robustness metric. The separation between the arithmetic mean and the -2/3rds mean of a set of probabilistic forecasts gives an indication of the degree of fluctuations about the central tendency, measured by the geometric or 0th mean.

Identification of a method for assessing probabilistic forecasts on the probability scale opens up other possibilities for integrating analysis with visual representations of performance. Recently, it was shown [18] that a calibration curve comparing reported probabilities and the measured distribution of the test samples can be overlaid with metrics using the generalized mean of the reported and measured probabilities. This approach uses the relationship that the probability associated with cross-entropy is the product of the probabilities associated with entropy and divergence (2.5) to distinguish between sources of uncertainty due to insufficient features versus insufficient models, respectively. As the utility of measuring the generalized mean of a set of probabilities is explored, further innovations can be developed for robust, accurate probabilistic forecasting. These are particularly important for the development of machine learning and artificial intelligence applications which need to carefully manage the uncertainty in making risky decisions.

## Appendix: Modeling risk as a coupling of statistical states

This chapter shows how risk sensitivity *r* can be used to evaluate the performance of probabilistic forecasts. In describing this assessment method an effort was made to keeping the explanation of the model as simple of possible. The purpose of this appendix is to expand upon the theoretical origins of the model. The model derives from the Tsallis generalization of statistical mechanics for complex systems [27]–[29] and utilizes a perspective based on the degree of nonlinear coupling between the statistical states of a system [12], [27].

The nonlinearity $\kappa$ of a complex systems increases the uncertainty about the long-range dynamics of the system. In [12] the effect of nonlinearity, such as multiplicative noise or variation in the variance, was shown to result in a modification from the exponential family $f(x) \propto e^{-x^\alpha}$ to the power-law domain with $\lim_{x \to \infty} f(x) \propto x^{-\frac{\alpha}{r}}$, where the risk sensitivity can be decomposed into the nonlinear coupling and the power and dimension of the variable, $r(\kappa, \alpha, d) = \frac{-\alpha\kappa}{1+d\kappa}$. As the source of coupling $\kappa$ increases from zero to infinity, the increased nonlinearity results in increasingly slow decay of the tails of the resulting distributions. Negative coupling can also be modeled, resulting in compact-support domain distributions with less variation than the exponential family. The negative domain, which models compact-support distributions, is $-\frac{1}{d} < \kappa < 0$.

The relationship defining *r* is also known within the field of nonextensive statistical mechanics as a dual transformation between the heavy-tail and compact support domains. With the alpha term dropping out the dual has the following relationship $\hat{\kappa} \Leftrightarrow \frac{-\kappa}{1+d\kappa}$. The dual is used to determine the conjugate to the decisive risk bias of 1. Taking $\alpha = 2$ and $d = 1$, the coupling for a risk bias of one is $1 = \frac{-2\kappa}{1+\kappa} \Rightarrow \kappa = -\frac{1}{3}$ and the conjugate values are $\hat{\kappa} = \frac{\frac{1}{3}}{1 - \frac{1}{3}} = \frac{1}{2}$ and $\hat{r} = \frac{-2\left(\frac{1}{2}\right)}{1 + \frac{1}{2}} = -\frac{2}{3}$.

The risk bias is closely related to the Tsallis entropy parameter $q = 1 - r$ [27]–[29]. One of the motivating principals of the Tsallis entropy methods was to examine how power law systems could be modeled using probabilities raised to a power $p_i^q$ [30]. As such, *q* can be thought of as the number of random variables needed to properly formulate the statistics of a complex system, while *r* represents the deviation from a linear system governed by exponential statistics. When the deformed probabilities are renormalized the resulting distribution can be shown to also represent the probability of a "coupled state" of the system

$$P_i^r = \frac{p_i^{1-r}}{\sum_{j=1}^n p_j^{1-r}} = \frac{p_i \prod_{\substack{k=1 \\ k \neq i}}^n p_k^r}{\sum_{j=1}^n \left( p_j \prod_{\substack{k=1 \\ k \neq j}}^n p_k^r \right)}, \tag{A.1}$$

hence use of the phrase "nonlinear statistical coupling".

Just as the probabilities are deformed via a multiplicative coupling, the entropy function is deformed via a nonlinear coupling term. The non-additivity of the generalized entropy $H_\kappa$ provides a definition for the degree of nonlinear coupling. The joint coupled-entropy of two independent systems *A* and *B* includes a nonlinear term

$$H_\kappa(A,B) = H_\kappa(A) + H_\kappa(B) + \kappa H_\kappa(A) H_\kappa(B). \tag{A.2}$$

For $\kappa = 0$ the additive property of entropy is satisfied by the logarithm of the probabilities. The function which satisfies the nonlinear properties of the generalized entropy is a generalization of the logarithm function referred to as the coupled logarithm

$$\ln_\kappa x \equiv \frac{1}{\kappa}\left( x^{\frac{\kappa}{1+\kappa}} - 1 \right), x > 0. \tag{A.3}$$

In the limit when $\kappa$ goes to zero the function converges to the natural logarithm. This definition of the generalized logarithm has the property that $\int_0^1 \ln_\kappa p^{-1} dp = 1$, thus the deformation modifies the relative information of a particular probability while preserving the 'total' information across the domain of probabilities.

The inverse of this function is the coupled exponential

$$\exp_\kappa x \equiv (1+\kappa x)^{\frac{1+\kappa}{\kappa}}. \tag{A.4}$$

A distribution of the exponential family will typically include an argument of the form $-x^\alpha/\alpha$ which is generalized by the relationship $\left(\exp_\kappa x^\alpha\right)^{\frac{-1}{\alpha}} = \exp_\kappa^{-1/\alpha} x^\alpha = (1+\kappa x^\alpha)^{\frac{1+\kappa}{-\alpha\kappa}}$. The rate of decay for a $d$-dimensional distribution is accounted for by $\exp_{\kappa,d}^{-1/\alpha} x^\alpha = (1+\kappa x^\alpha)^{\frac{1+d\kappa}{-\alpha\kappa}}$, neglecting the specifics of the matrix argument. This is the form of the multivariate Student's t distribution with $\kappa$ equal to the inverse of the degree of freedom. When the generalized logarithm needs to include the role of the power and dimension this is expressed as $\ln_{\kappa,d} x^{-\alpha} \equiv \frac{1}{\kappa}\left(x^{\frac{-\alpha\kappa}{1+d\kappa}} - 1\right)$ or alternatively $\left(\ln_{\kappa,d} x^{-\alpha}\right)^{\frac{1}{\alpha}} = \left(\frac{1}{\kappa}\left(x^{\frac{-\alpha\kappa}{1+d\kappa}} - 1\right)\right)^{\frac{1}{\alpha}}$. The first expression is used here, though research regarding the later expression has been explored.

There are a variety of expressions for a generalized entropy function which when translated back to the probability domain lead to the generalized mean of a probability distribution. Generalization of the entropy function can be viewed broadly as a modification of the logarithm function and the weight of the arithmetic mean. The translation back to the probability domain makes use of the inverse of the generalized logarithm, namely the generalized exponential. The generalized expression for aggregating probabilities is then

$$\begin{aligned}
P_\kappa(\mathbf{w},\mathbf{p};\alpha,d) &= \exp_{\kappa,d}^{-1/\alpha}\left(H_\kappa(\mathbf{p};\mathbf{w},\alpha,d)\right) \\
&= \exp_{\kappa,d}^{-1/\alpha}\left(\sum_{i=1}^N w_i \ln_{\kappa,d} p_i^{-\alpha}\right) \\
&= \left(1+\kappa\left(\sum_{i=1}^N \frac{w_i}{\kappa}\left(p_i^{\frac{-\alpha\kappa}{1+d\kappa}} - 1\right)\right)\right)^{\frac{1+d\kappa}{-\alpha\kappa}} \\
&= \left(\sum_{i=1}^N w_i p_i^{\frac{-\alpha\kappa}{1+\kappa}}\right)^{\frac{1+d\kappa}{-\alpha\kappa}},
\end{aligned} \tag{A.5}$$

where the weights $w_i$ are assumed to sum to one. In the main text the focus is placed on the risk bias $r = \dfrac{-\alpha\kappa}{1+d\kappa}$, which forms the power of the generalized mean. The coupled entropy function is defined using the coupled probability (A.1) for the weights. Other generalized entropy functions use different definitions for the weights and generalized logarithm, but as proven in [12] for at least the normalized Tsallis entropy, Tsallis entropy and Rényi entropy they all converge to the weighted generalized mean of the distribution

$$P_\kappa(\mathbf{p};\alpha,d) = \left( \sum_{i=1}^N \left( \frac{p_i^{1+\frac{\alpha\kappa}{1+d\kappa}}}{\sum_{j=1}^N p_j^{1+\frac{\alpha\kappa}{1+d\kappa}}} \right) p_i^{\frac{-\alpha\kappa}{1+d\kappa}} \right)^{\frac{1+d\kappa}{-\alpha\kappa}}$$

$$= \left( \frac{\sum_{i=1}^N p_i}{\sum_{j=1}^N p_j^{1+\frac{\alpha\kappa}{1+d\kappa}}} \right)^{\frac{1+d\kappa}{-\alpha\kappa}} \tag{A.6}$$

$$= \left( \sum_{i=1}^N p_i^{1+\frac{\alpha\kappa}{1+d\kappa}} \right)^{\frac{1+d\kappa}{\alpha\kappa}}.$$

The assessment of a probabilistic forecast treats each test sample as an independent equally likely event. The weights, even using the coupled probability, simplify to one over the number of test samples

$$w_i = \frac{\left(1/N\right)^{1+\frac{\alpha\kappa}{1+d\kappa}}}{\sum_{i=1}^N \left(1/N\right)^{1+\frac{\alpha\kappa}{1+d\kappa}}} = \frac{\left(1/N\right)^{1+\frac{\alpha\kappa}{1+d\kappa}}}{N\left(\left(1/N\right)^{1+\frac{\alpha\kappa}{1+d\kappa}}\right)} = \frac{1}{N}. \tag{A.7}$$

Thus the generalized mean used for the Risk Profile has a power with the opposite sign of that used for the average probability of distribution

$$P_\kappa(\mathbf{p};\alpha,d) = \left( \frac{1}{N} \sum_{i=1}^N p_i^{\frac{-\alpha\kappa}{1+d\kappa}} \right)^{\frac{1+d\kappa}{-\alpha\kappa}}, \tag{A.8}$$

where the probabilities in this express are samples from a set of forecasted of events.

## Acknowledgement

The initial research developing the *Risk Profile* was sponsored by Raytheon IRAD IDS202 2010-2012. Conversations with Ethan Phelps and Herb Landau highlighted the need for the surprisal metric in the assessment of discrimination algorithms. The figures in section 4 were created using Matlab software developed by Brian Scannell.